\documentclass[twocolumn,showpacs,preprintnumbers,amsmath,amssymb,superscriptaddress,nofootinbib]{revtex4-1} 
\usepackage{mathrsfs, epsfig, graphicx, url, dcolumn, bm,natbib}
\usepackage{hyperref}
\usepackage{hypernat}
\usepackage{aas_macros}

\begin{document}

\title{The Linear Perturbation Theory of Reionization in Position-Space: Cosmological Radiative Transfer Along the Light-Cone}

\author{Yi Mao}
\email{mao@iap.fr}
\affiliation{Institut d'Astrophysique de Paris, Institut Lagrange de Paris, CNRS, UPMC Univ Paris 06, UMR7095, 98 bis, boulevard Arago, F-75014, Paris, France}
\author{Anson D'Aloisio}
\affiliation{Department of Astronomy, University of Washington, Box 351580, Seattle, WA, 98195}
\author{Benjamin D. Wandelt}
\affiliation{Institut d'Astrophysique de Paris, Institut Lagrange de Paris, CNRS, UPMC Univ Paris 06, UMR7095, 98 bis, boulevard Arago, F-75014, Paris, France}
\affiliation{Department of Physics, University of Illinois at Urbana-Champaign, Urbana, IL 61801}
\author{Jun Zhang}
\affiliation{Center for Astronomy and Astrophysics, Department of Physics and Astronomy, Shanghai Jiao Tong University, 955 Jianchuan road, Shanghai, 200240, China}
\author{Paul R. Shapiro}
\affiliation{Department of Astronomy and Texas Cosmology Center, University of Texas, Austin, Texas 78712}

\date{submitted 28 November, 2014; accepted 13 April, 2015}

\begin{abstract}

The linear perturbation theory of inhomogeneous reionization (LPTR) has been developed as an analytical tool for predicting the global ionized fraction and large-scale power spectrum of ionized density fluctuations during reionization.  In the original formulation of the LPTR, the ionization balance and radiative transfer equations are linearized and solved in Fourier space.  However, the LPTR's approximation to the full solution of the radiative transfer equation is not straightforward to interpret, since the latter is most intuitively conceptualized in position space.  To bridge the gap between the LPTR and the language of numerical radiative transfer, we present a new, equivalent, position-space formulation of the LPTR that clarifies the approximations it makes and facilitates its interpretation.  We offer a comparison between the LPTR and the excursion-set model of reionization (ESMR), and demonstrate the built-in capability of the LPTR to explore a wide range of reionization scenarios, and to go beyond the ESMR in exploring scenarios involving X-rays. 

\end{abstract}


\setcounter{footnote}{0}


\maketitle

\section{Introduction}

The epoch of reionization (EOR) is at the frontier of observational cosmology. Analytical models of the EOR have played an important role in our theoretical understanding of reionization because they provide a readily accessible and computationally inexpensive (albeit approximate) method of exploring salient features of the EOR.  An interesting example is the linear perturbation theory of inhomogeneous reionization (LPTR), originally formulated in Ref.~\cite{2007MNRAS.375..324Z}, in which the equations of radiative transfer and ionization balance are expanded to first order in perturbations to the ionizing radiation and H~I density fields. Ref.~\cite{2007MNRAS.375..324Z} showed that there is an exact solution to the linearized radiative transfer equation in {\it Fourier} space, and developed analytical machinery for calculating the global ionized fraction and large-scale clustering of ionized hydrogen density during the EOR. 

The LPTR has since shown its applicability to a variety of problems. For example, using the LPTR, Ref.~\cite{2010ApJ...711.1310K} computed the cosmic microwave background bispectrum due to inhomogeneous reionization; Ref.~\cite{2013MNRAS.433.2900D} investigated the effect of primordial non-Gaussianity on the clustering of ionized density during the EOR, and their results were then applied in Ref.~\cite{2013PhRvD..88h1303M} to forecast the detectability of primordial non-Gaussianity using the EOR 21~cm power spectrum. Ref.~\cite{2014PhRvD..89h3010P} performed an independent calculation in a similar spirit to the LPTR to study neutral density fluctuations of the intergalactic medium (IGM) in the post-reionization epoch. 

Formulating the LPTR in Fourier space is a matter of mathematical convenience.  However, the full solution of the radiative transfer equation is most intuitively expressed in position space. This disparity between the LPTR and the language of numerical radiative transfer makes interpretation of the former, as well as the approximations it entails, difficult.   Bridging this gap requires a reformulation of the LPTR in which approximations to full radiative transfer are made explicitly in position space.  In this paper, we present such a position-space reformulation in order to lay a clearer theoretical foundation for the LPTR, which may help enlarge its regime of application. 

Another useful analytical model of reionization --- the excursion set model of reionization (ESMR) -- was originally proposed in Ref.~\cite{2004ApJ...613....1F} and subsequently developed as the basis of several ``semi-numerical" simulation codes \cite{2009MNRAS.394..960C,2010MNRAS.406.2421S,2011MNRAS.411..955M,2014Natur.506..197F}.  These ESMR-based semi-numerical algorithms have been shown to agree well with full radiative transfer simulations for the simplest models of reionization \cite{2011MNRAS.414..727Z,2014MNRAS.443.2843M}. We shall borrow the analytical ESMR that inspired these algorithms (i.e \emph{not} the semi-numerical algorithms themselves) as a reference for comparison with the LPTR, not only to cross-check their consistency under similar assumptions, but to identify the scenarios that the LPTR can explore beyond the limits of the ESMR. 

The rest of this paper is organized as follows. In Section~\ref{sec:reformulation}, we summarize and re-express the LPTR in terms of the conventional variables and language of radiative transfer. In Section~\ref{sec:rederivation}, we reformulate the LPTR in position space. With the insights provided by our reformulation, we discuss the qualitative differences between the LPTR and ESMR in Section~\ref{sec:comparison_qualitative}, and we present numerical results for a range of simple reionization models in Section~\ref{sec:numerical_results}. Finally, we offer concluding remarks in Section~\ref{sec:conclusion}. 

\section{The LPTR re-expressed}
\label{sec:reformulation}

The original LPTR formalism in Ref.~\cite{2007MNRAS.375..324Z} was expressed using some non-standard notation and variables. We begin here by re-expressing the LPTR in the standard language of radiative transfer in order to facilitate the LPTR's physical interpretation. 

Consider a ray with the comoving specific intensity $I_\nu (\eta, {\bf x},\nu,{\bf n})$ at the conformal time $\eta$ on comoving coordinates ${\bf x}$ with the proper frequency $\nu$, 
which is the flux of energy received by an observer in the comoving frame, $a\,h_{\rm pl}\nu dN$ (where $a$ is the cosmic scale factor, $h_{\rm pl}$ is the Planck constant, and $dN$ is the number of photons), per unit conformal time $d\eta = dt/a$, per unit comoving receiving area $dA_{\perp}/a^2$, per unit comoving frequency interval $a\,d\nu$, from unit solid angle $d{\bf n}$ about the transport direction ${\bf n}$ (unit vector). The comoving specific intensity is related to the proper one by $I_\nu = a^3 I_{\nu,{\rm proper}}$.
The radiative transfer equation in the Eulerian scheme is 
\begin{equation}
\frac{\partial I_\nu}{c\partial \eta} + {\bf n}\cdot\nabla I_\nu - \frac{aH\nu}{c} \frac{\partial I_\nu}{\partial \nu} = S_\nu - \Gamma_\nu I_\nu\,.
\label{eqn:RTeqn1}
\end{equation}
Here $S_\nu(\eta, {\bf x},\nu,{\bf n})$ is the comoving source emissivity, and $\Gamma_\nu(\eta, {\bf x},\nu)$ is the comoving absorption coefficient, and $H(a)$ is the Hubble parameter. The comoving quantities are related to their proper counterparts by $S_\nu = a^4 j_\nu$, and $\Gamma_\nu = a \kappa_\nu$, respectively. In the case of photoionization of hydrogen atoms, for example, $\Gamma_\nu(\eta, {\bf x},\nu) = \sigma_{\rm HI}(\nu) n_{\rm HI}(\eta, {\bf x})/a^2$, where $\sigma_{\rm HI}(\nu)$ is the cross-section of hydrogen photoionization at the proper frequency $\nu$ and $n_{\rm HI} = n_{\rm HI, proper}\,a^3 $ is the comoving number density of H~I atoms. 

Now we develop a trick that was introduced in Ref.~\cite{2007MNRAS.375..324Z}, which can formally remove the partial differentiation with respect to frequency.  We define the new time and frequency variables $\xi = \ln a$ and $\zeta = \ln(\nu/\nu_{\rm H})$ (where $\nu_{\rm H}$ is the hydrogen Lyman-limit frequency). So, $d\xi = aHd\eta$ and $d\zeta = d\nu/\nu$. We further define another new set of mixed time-frequency variables $u = \frac{1}{2}(\xi + \zeta)$ and $v = \frac{1}{2}(\xi - \zeta)$, or $\xi = u + v$ and $\zeta = u - v$, so $(\partial/\partial v)|_u = (\partial/\partial\xi)|_\zeta - (\partial/\partial\zeta)|_\xi$. It is straightforward to show that eq.~(\ref{eqn:RTeqn1}) can be rewritten as 
$\left.(aH/c)(\partial I_\nu/\partial v)\right|_u + {\bf n}\cdot\nabla I_\nu =  S_\nu - \Gamma_\nu I_\nu $.

Now we define the comoving\footnote{The idea of using the comoving frequency to simplify the radiative transfer equation was applied earlier in Ref.~\cite{1996ApJS..102..191G}.} frequency $f= \nu a$. 
Note that $u = \frac{1}{2}\ln(f/\nu_{\rm H})$, so fixing $u$ is equivalent to fixing $f$, which accounts for cosmological redshifting. When $f$ is fixed, $dv = d\xi = aHd\eta$, so eq.~(\ref{eqn:RTeqn1}) can be rewritten as
\begin{equation}
\left.\frac{\partial I_\nu}{c\partial \eta}\right|_{f} + {\bf n}\cdot\nabla I_\nu =  S_\nu  - \Gamma_\nu I_\nu \,.
\label{eqn:RTeqn3}
\end{equation}
The LHS of eq.~(\ref{eqn:RTeqn3}) does not contain the derivative with respect to frequency $f$. 
Hereafter we relabel the dependence of all fields on $\nu$ by their dependence on $f$, e.g.\ $I_\nu (\eta, {\bf x},f,{\bf n})$. 

\subsection{Spatial averages}

We define the spatial average of the specific intensity, the source emissivity, and the intensity-weighted mean absorption coefficient as follows, respectively. 
\begin{eqnarray}
\overline{I}_\nu (\eta,f) &=& \frac{1}{4\pi V}\int d^3 x \int d^{2}{\bf n}\,\, I_\nu(\eta,{\bf x},f,{\bf n})\,,\\
\overline{S}_\nu (\eta,f) &=& \frac{1}{4\pi V}\int d^3 x \int d^{2}{\bf n}\,\, S_\nu(\eta,{\bf x},f,{\bf n})\,,\\
\overline{\Gamma}_{\nu,{\rm I}} (\eta,f) &=& \overline{\Gamma_\nu I_\nu}(\eta,f)/\overline{I_\nu}(\eta,f) \,,
\end{eqnarray}
where 
\begin{equation}
\overline{\Gamma_\nu I_\nu} (\eta,f) = \frac{1}{4\pi V}\int d^3 x \int d^{2}{\bf n} \Gamma_\nu(\eta,{\bf x},f) I_\nu(\eta,{\bf x},f,{\bf n})\,.
\end{equation}
Here $V$ is the total volume. 

From eq.~(\ref{eqn:RTeqn3}), $\overline{I}_\nu$ is governed by 
\begin{equation}
\left.\frac{\partial \overline{I}_\nu}{c\partial \eta}\right|_{f} =  \overline{S}_\nu - \overline{\Gamma}_{\nu,{\rm I}} \,\overline{I}_\nu \,.
\label{eqn:RTeqn4}
\end{equation}
The exact solution to eq.~(\ref{eqn:RTeqn4}) is 
\begin{equation}
\overline{I}_\nu(\eta,f) = \int_{\eta_0}^{\eta} c d\eta_s \exp[-\overline{\tau}_{\rm I}(\eta,\eta_s;f)]\,\overline{S}_\nu(\eta_s,f) \,.
\label{eqn:keyeq_globalmean}
\end{equation}
Throughout this paper, the initial time $\eta_0$ is chosen to be early enough that there is no ionizing radiation. 
We define the intensity-weighted mean optical depth for the time interval between $\eta$ and $\eta_s$ 
\begin{equation}
 \overline{\tau}_{\rm I}(\eta,\eta_s;f) \equiv \int_{\eta_s}^{\eta} c d\eta' \,\overline{\Gamma}_{\nu,{\rm I}}(\eta',f)\,.
\end{equation}
Eq.~(\ref{eqn:keyeq_globalmean}) was derived earlier in Refs.~\cite{Shapiro:1993hn,1996ApJS..102..191G}, and
is equivalent to the original LPTR expression, eq.~(17) in Ref.~\cite{2007MNRAS.375..324Z}. 
Note that the use of radiation-H~I clumping factor $C_{\gamma\,H}^{(2)}$ in Ref.~\cite{2007MNRAS.375..324Z} is implicitly included in our definition of the intensity-weighted mean absorption coefficient herein. 

\subsection{The linear perturbations}

Now we write the perturbations in the fields 
\begin{eqnarray}
\Delta I_\nu (\eta,{\bf x},f,{\bf n}) &=& I_\nu (\eta,{\bf x},f,{\bf n}) - \overline{I}_\nu (\eta,f)\,,\\
\Delta S_\nu (\eta,{\bf x},f,{\bf n}) &=& S_\nu (\eta,{\bf x},f,{\bf n}) - \overline{S}_\nu (\eta,f)\,,\\
\Delta \Gamma_{\nu,{\rm I}} (\eta,{\bf x},f) &=& \Gamma_\nu (\eta,{\bf x},f) - \overline{\Gamma}_{\nu,{\rm I}} (\eta,f)\,, \\
\Delta \Gamma_\nu (\eta,{\bf x},f) &=& \Gamma_\nu (\eta,{\bf x},f) - \overline{\Gamma}_\nu (\eta,f)\,,
\end{eqnarray}
where $\overline{\Gamma}_\nu$ is the volume-weighted mean absorption coefficient,
\begin{equation}
\overline{\Gamma}_\nu (\eta,f) = \frac{1}{V}\int d^3 x \Gamma_\nu(\eta,{\bf x},f)\,.
\end{equation}

From eq.~(\ref{eqn:RTeqn3}), the perturbations at a given $f$ and ${\bf n}$ are governed by
\begin{equation}
\left.\frac{\partial \Delta I_\nu}{c\partial \eta}\right|_{f} + {\bf n}\cdot\nabla (\Delta I_\nu) =  \Delta S_{\nu,{\rm eff}}   - \overline{\Gamma}_\nu \Delta I_\nu - \Delta \Gamma_\nu \Delta I_\nu \,,
\label{eqn:RTeqn5}
\end{equation}
where the {\it effective} perturbation in source emissivity is defined as 
\begin{equation}
\Delta S_{\nu,{\rm eff}} = \Delta S_\nu - \overline{I}_\nu \Delta \Gamma_{\nu,{\rm I}}\,.
\end{equation}
The effective perturbation in source emissivity is the perturbation in source emissivity minus the perturbation in the photoionization rate in the {\it mean} radiation field. 

Now, the radiative transfer equation is linearized, i.e.\ to first order in perturbations, so the $\Delta \Gamma_\nu \Delta I_\nu$ term is dropped from the RHS of eq.~(\ref{eqn:RTeqn5}). Then we Fourier transform \footnote{We use an overhead tilde, e.g.\  $\tilde{f}({\bf k})$, to denote the Fourier transform of some field $f({\bf x})$ throughout this paper.} the linearized radiative transfer equation, 
\begin{equation}
\left.\frac{\partial \widetilde{\Delta I_\nu}({\bf k})}{c\partial \eta}\right|_{f} =  \widetilde{\Delta S}_{\nu,{\rm eff}}({\bf k})  - \left(i{\bf n}\cdot{\bf k}  + \overline{\Gamma}_\nu \right) \widetilde{\Delta I_\nu}({\bf k})  \,,
\label{eqn:RTeqn6}
\end{equation}
where the dependences on $\eta$, $f$ and ${\bf n}$ are implicit. The exact solution is 
\begin{eqnarray}
& & \widetilde{\Delta I_\nu}(\eta,{\bf k},f,{\bf n}) = \int_{\eta_0}^{\eta} c\,d\eta_s \,\exp[-\overline{\tau}(\eta,\eta_s;f)]\,\nonumber \\ 
& & \qquad\times \exp\Bigl[-i{\bf n}\cdot{\bf k}\chi(\eta,\eta_s)\Bigr]\, \widetilde{\Delta S}_{\nu,{\rm eff}}(\eta_s,{\bf k},f,{\bf n})\,,
\label{eqn:RTeqn6.1}
\end{eqnarray}
where $\chi(\eta,\eta_s) \equiv \int_{\eta_s}^{\eta} c d\eta'$ is the comoving line-of-sight distance for a time interval, and we define $\overline{\tau}$ as the volume-weighted mean optical depth,
\begin{equation}
 \overline{\tau}(\eta,\eta_s;f) \equiv \int_{\eta_s}^{\eta} c d\eta' \overline{\Gamma}_{\nu}(\eta',f)\,.
\label{eqn:meantau}
\end{equation}

For atoms at a given location, the photoionization rate only depends on the sum of radiation intensity over all directions, i.e.\ the monopole intensity $\int d^2{\bf n}\,I_\nu = 4\pi\,\overline{I}_\nu + \int d^2{\bf n}\,\Delta I_\nu$. The second term on the RHS here is the Fourier transform of $\int d^2{\bf n}\,\widetilde{\Delta I_\nu}$ that can be calculated by analytically integrating the solution for $\widetilde{\Delta I_\nu}$ over solid angle. We further assume that the source emissivity is isotropic, i.e.\ $S_{\nu} = S_{\nu}(\eta_s,{\bf x},f)$, which is approximately valid for ionizing sources on large scales. In this case, the solution for the perturbation in the monopole intensity in Fourier space is 
\begin{eqnarray}
& & \int d^2{\bf n}\,\widetilde{\Delta I_\nu}(\eta,{\bf k},f,{\bf n}) = 4\pi\,\int_{\eta_0}^{\eta} c\,d\eta_s \,\exp[-\overline{\tau}(\eta,\eta_s;f)]\,\nonumber \\ 
& & \qquad\times j_0\Bigl(\chi(\eta,\eta_s)\left|{\bf k}\right|\Bigr)\, \widetilde{\Delta S}_{\nu,{\rm eff}}(\eta_s,{\bf k},f)\,,
\label{eqn:keqeq_pertub_Fourier}
\end{eqnarray}
where $j_0(x)=\sin(x)/x$ is the spherical Bessel function of the first kind. 
It is straightforward to check that eq.~(\ref{eqn:keqeq_pertub_Fourier}) is equivalent to the complicated expression in the original LPTR, eq.~24 in Ref.~\cite{2007MNRAS.375..324Z}. 

The LPTR solution (eq.~\ref{eqn:keqeq_pertub_Fourier}) is interpreted as follows. The effective perturbation in  source emissivity propagates in Fourier space to the perturbation in monopole intensity along the light-cone, subject to two effects --- the attenuation by the mean optical depth, and the modulation by the scale-dependent factor $j_0(\chi\left|{\bf k}\right|)$. While the Fourier space formulation is mathematically elegant, this interpretation is somehow opaque, because the solution to the full radiative transfer equation is conceptually intuitive in position space. In particular, it is not clear what approximations made in the LPTR result in each above effect separately. In what follows, we reformulate the LPTR in position space to address this question. 

\section{Reformulation in position space}
\label{sec:rederivation}

We start from the full radiative transfer equation for the perturbation in specific intensity, $\Delta I_\nu$, in position space, i.e.\ eq.~(\ref{eqn:RTeqn5}) written in the Eulerian scheme.  It can be rewritten in the Lagrangian scheme as 
\begin{equation}
\frac{d \Delta I_\nu}{ds} =  \Delta S_{\nu,{\rm eff}}   - \Gamma_\nu \Delta I_\nu \,,
\label{eqn:RTeqn7}
\end{equation}
where $d/ds$ is the total derivative along the (comoving) ray path, $ds = c\,d\eta$. The solution to the full radiative transfer equation in the Lagrangian scheme is well-known, 
\begin{equation}
\Delta I_\nu(s) = \int ds' \Delta S_{\nu,{\rm eff}}(s') \exp\left[-\int_{s'}^{s} ds'' \Gamma_{\nu}(s'')\right]\,.
\end{equation}
Here we assume again that at an early enough time, there is no ionizing radiation. 
This solution to the full radiative transfer equation can be rewritten back to the Eulerian scheme, as 
\begin{eqnarray}
& & \Delta I_\nu(\eta,{\bf x},f,{\bf n}) = \int_{\eta_0}^{\eta} c\,d\eta_s\,\Delta S_{\nu,{\rm eff}}(\eta_s,{\bf x}-\chi(\eta,\eta_s){\bf n},f,{\bf n}) \nonumber\\ 
& & \qquad\times \exp\left[-\int_{\eta_s}^{\eta} c\,d\eta' \Gamma_{\nu}(\eta',{\bf x}-\chi(\eta,\eta'){\bf n},f)\right]\,.
\label{eqn:RTeqn8}
\end{eqnarray}
Note that the emission events at $(\eta_s,{\bf x}-\chi(\eta,\eta_s){\bf n})$ and photoionization events at $(\eta',{\bf x}-\chi(\eta,\eta'){\bf n})$ are on the past light-cone of the observation event at $(\eta,{\bf x})$.

Now, in the spirit of linear perturbation theory, since both $\Delta I_\nu$ and $\Delta S_{\nu,{\rm eff}}$ are perturbations, we can neglect the perturbations in the optical depth, i.e.\ replace $\int_{\eta_s}^{\eta} c\,d\eta' \Gamma_{\nu}(\eta',{\bf x}-\chi(\eta,\eta'){\bf n},f)$ by $\int_{\eta_s}^{\eta} c\,d\eta' \overline{\Gamma}_{\nu}(\eta',f) = \overline{\tau}(\eta,\eta_s;f)$ which is independent of position and propagation direction. It is important to note that the mean optical depth is not the same as the mean optical depth that would be calculated through the universe in the limit of infinite speed of light, i.e.\ for which the opacity of the universe along the line of travel of emitted photons is evaluated at a single cosmic time, the time at which the local intensity is calculated. In fact, our mean optical depth is not only integrated along the past light-cone, but can be thought of as the average of the optical depth $\int_{\eta_s}^{\eta} c\,d\eta' \Gamma_{\nu}(\eta',{\bf x}-\chi(\eta,\eta'){\bf n},f)$ of each individual ray, over different arrival directions for photons reaching the point of observation, and over all such observation points. 

In addition, we assume that the source emissivity is isotropic, so the explicit dependence of $\Delta S_{\nu,{\rm eff}}$ on the propagation direction ${\bf n}$ is removed. However, the source emissivity is still implicitly dependent on ${\bf n}$ through the explicit dependence on position. With these two assumptions, the solution to the radiative transfer equation is simplified as 
\begin{eqnarray}
& & \Delta I_\nu(\eta,{\bf x},f,{\bf n}) = \int_{\eta_0}^{\eta} c\,d\eta_s\,\exp\left[-\overline{\tau}(\eta,\eta_s;f)\right]\nonumber \\
& & \qquad\times \Delta S_{\nu,{\rm eff}}(\eta_s,{\bf x}-\chi(\eta,\eta_s){\bf n},f) \,.
\end{eqnarray}

Integrating over ${\bf n}$, the perturbation in monopole intensity at a given location is 
\begin{eqnarray}
& & \int d^2{\bf n}\,\Delta I_\nu(\eta,{\bf x},f,{\bf n}) = \int_{\eta_0}^{\eta} c\,d\eta_s\,\exp\left[-\overline{\tau}(\eta,\eta_s;f)\right]\nonumber \\
& & \qquad\times \int d^2{\bf n} \,\Delta S_{\nu,{\rm eff}}(\eta_s,{\bf x}-\chi(\eta,\eta_s){\bf n},f) \,.
\label{eqn:RTeqn9}
\end{eqnarray}
Let us focus on the second line in eq.~(\ref{eqn:RTeqn9}), the integration of $\Delta S_{\nu,{\rm eff}}$ on a spherical surface of radius $\chi(\eta,\eta_s)$ with center at ${\bf x}$. It is easy to show that this integration is equal to 
\begin{equation}
\int d^3 x' \frac{1}{\chi^2(\eta,\eta_s)} \delta\Bigl(\left|{\bf x}-{\bf x'}\right|-\chi(\eta,\eta_s)\Bigr)\,\Delta S_{\nu,{\rm eff}}(\eta_s,{\bf x'},f)\,.
\label{eqn:convol}
\end{equation}  
Here $\delta(r)$ is the 1D Dirac-$\delta$ function. 

Eq.~(\ref{eqn:convol}) is {\it de facto} the convolution of two fields, $\frac{1}{\chi^2(\eta,\eta_s)} \delta\Bigl(\left|{\bf x}\right|-\chi(\eta,\eta_s)\Bigr)$, and $\Delta S_{\nu,{\rm eff}}(\eta_s,{\bf x},f)$. The Fourier transform of the convolution is the product of the Fourier transform of the separate fields. Since the Fourier transform of $\frac{1}{\chi^2} \delta(\left|{\bf x}\right|-\chi)$ is $4\pi\,j_0(\chi|{\bf k}|)$, the Fourier transform of the second line in eq.~(\ref{eqn:RTeqn9}) is 
\begin{equation}
4\pi\,j_0\Bigl(\chi(\eta,\eta_s)\left|{\bf k}\right|\Bigr)\, \widetilde{\Delta S}_{\nu,{\rm eff}}(\eta_s,{\bf k},f)\,.
\end{equation} 
With this help, Fourier transforming eq.~(\ref{eqn:RTeqn9}), we recover exactly the LPTR solution of perturbations in Fourier space (eq.~\ref{eqn:keqeq_pertub_Fourier}). 

The physical interpretation of the LPTR solution is made clearer by this position-space reformulation. The effective perturbation in source emissivity propagates in position space to the perturbation in specific intensity along the light-cone, subject to two aforementioned effects: (1) the attenuation by the mean optical depth, resulting from the neglect of inhomogeneous photoionization rate along light rays; (2) the modulation by the $j_0(\chi |{\bf k}|)$ factor, representing the integration of the effective perturbation in isotropic source emissivity over different photon arrival directions from a 2D spherical surface of equal look-back time along the past light-cone. 
Note that the LPTR formalism takes into account the spatial fluctuations of all physical quantities to first order. For example, the fluctuation of the photoionization rate is included through the $\overline{I}_\nu \Delta \Gamma_{\nu,{\rm I}}$ and $\Delta I_\nu \overline{\Gamma}_{\nu}$ term. The neglect of the fluctuations of optical depth in LPTR, nevertheless, may underestimate the fluctuations of photon intensity, and, hence, suppress the fluctuations of ionized fraction on small scales.

This reformulation also makes clearer what contributes to the inhomogeneity in the radiation field and H~II distributions in the LPTR. Since the attenuation is accounted for by the \emph{mean} optical depth, the inhomogeneity in the radiation field results from the variations in the integrated effective perturbations in source emissivity. However, in the extremely hard spectrum case in which the photon mean-free-path is very large, the integration on a large spherical surface smoothes out the effective perturbations in source emissivity, regardless of the location of the center. In this case, the radiation intensity approaches to the homogeneous case, as considered in Ref.~\cite{Shapiro:1993hn,1996ApJS..102..191G}. On the other hand, the inhomogeneity in the H~II distribution results from not only the inhomogeneity in the radiation field, but the fluctuations in hydrogen recombination rate, the latter of which depends on the local overdensity of hydrogen atoms. 

\begin{table*}
\caption{Illustrative models of reionization. As an example of the naming convention for the LPTR models, the model labelled as ``LPTR-2As'' corresponds to $C_{\rm HII}=2$, source Model A, and a soft spectrum with $s=-3$. The time-dependence of $C_{\rm HII}$ refers to the variation with redshift according to the fitting formula in eq.~(\ref{eqn:time-varying-clumping}). In the LPTR, source Model A is the case in which the emissivity is proportional to the time derivative of the collapse fraction (eq.~\ref{eqn:sourceA}), while Model B is the case in which the emissivity is proportional to the collapse fraction (eq.~\ref{eqn:sourceB}). The source spectrum is parametrized according to eq.~(\ref{eqn:softness}), and the cases with the index $s=-3$ and $s=-1$ are representative examples of the soft and hard spectra, respectively. The simplest variant of the ESMR \cite{2004ApJ...613....1F} is limited to an effectively constant and homogeneous recombination rate, and consistent with the assumptions of source Model A with a soft spectrum. For each LPTR and ESMR model, the value of the efficiency parameter is determined by fixing $\tau_{\rm es} = 0.08$.
\label{tab:models}}
\begin{ruledtabular}
\begin{tabular}{ccccccl}

Model		& $C_{\rm HII}$  &  $C^{(1)}_{\gamma {\rm H}}$   &  $C^{(2)}_{\gamma {\rm H}}$   &	Emissivity model   &	Emissivity softness $s$  &  Efficiency parameter \\ \hline
ESMR 	& --- 	 &  ---  &  ---  &  ---    &  ---    &  $\zeta_{\rm ESMR}=50.2$ \\
LPTR-2As	& 2		 &  1	 &  1	 &  A			   &  -3		     &  $\zeta_{\rm LPTR}^{\rm A}= 54.2$ \\
LPTR-2Bs	& 2		 &  1	&  1	 &  B			   &  -3		     &  $\zeta_{\rm LPTR}^{\rm B}= 9.9\times 10^3$ \\
LPTR-tAs 	& time-dependent &  1	&  1	 &  A			   &  -3		     &  $\zeta_{\rm LPTR}^{\rm A}= 70.3$ \\
LPTR-tBs 	& time-dependent &  1	&  1	 &  B			   &  -3		     &  $\zeta_{\rm LPTR}^{\rm B}=1.18\times 10^4$ \\
LPTR-2Ah 	& 2 		 &  1 	&  1	 &  A			   &  -1		     &  $\zeta_{\rm LPTR}^{\rm A}=62$ \\
LPTR-2Bh	& 2 		 &  1 	&  1	 &  B			   &  -1		     &  $\zeta_{\rm LPTR}^{\rm B}=1.16\times 10^4$ \\
\end{tabular}
\end{ruledtabular}
\end{table*}

\section{Comparison with the ESMR}
\label{sec:comparison_qualitative}

Now that we have reformulated the LPTR in a way that makes clearer its approximations to radiative transfer, we offer a comparison between the LPTR and the ESMR of Ref. \cite{2004ApJ...613....1F}.  The comparison presented here is limited to the simplest variant of the ESMR, as originally developed by Ref. \cite{2004ApJ...613....1F}.  Though the ESMR has since been extended in several ways (see e.g. \cite{2005MNRAS.363.1031F,2007MNRAS.374...72C,2013ApJ...771...35K,2014ApJ...787..146K,2014MNRAS.442.1470P, 2014ApJ...781...97X}), we focus our discussion on the original/simplest version without loss of generality because the underlying principles of its more sophisticated extensions remain essentially the same.     

The basic assumptions of the ESMR can be summarized as follows: (1) Each galaxy of mass $M_\mathrm{gal}$ can ionize a mass $M_{\mathrm{ion}} = \xi_{\rm ESMR} M_{\mathrm{gal}}$, where $\xi_{\rm ESMR}$ is an efficiency parameter that depends on, for example, the star formation efficiency, the number ionizing photons produced per stellar baryon, the escape fraction of ionizing photons, and the mean recombination rate of the ionized IGM.  Due to the high degree of uncertainty in all of these parameters, $\xi_{\rm ESMR}$ is treated as a free parameter; (2)   A point $\boldsymbol{x}$ in the IGM is assumed to be part of an ionized region once a sphere centered on it contains enough mass in collapsed halos to ionize all of the neutral hydrogen atoms in that sphere.  Let $f_{\rm coll}(\eta,M_{\rm min},R,\delta_R)$ be the collapsed fraction of halos with masses above some threshold\footnote{For simplicity, we assume that halos with virial temperatures $T_{\rm vir}\ge 10^4\,{\rm K}$,  within which gas could be radiatively cooled through collisional excitation of atomic hydrogen, were the only sources of reionization. The minimum $T_{\rm vir}\ge 10^4\,{\rm K}$ criterion roughly corresponds to a minimum halo mass scale $M_{\rm min}\sim 10^8\,M_\odot$.} $M_{\rm min}$, where $\delta_R$ is the linearly extrapolated density contrast smoothed on scale $R$.  The condition for our fiducial point to be ionized is

\begin{equation}
 f_{\rm coll}(\eta,M_{\rm min},R,\delta_R) \ge \xi_{\rm ESMR}^{-1} 
 \label{EQ:ESMRcondition}
\end{equation}
More precisely, the point is assumed to be in an ionized region of mass $M=\bar{\rho}_m 4 \pi R^3 / 3$ (where $\bar{\rho}_m$ is the comoving mean matter density) for the \emph{largest} such sphere that satisfies the condition in eq.~(\ref{EQ:ESMRcondition}).  If this condition is not satisfied for any mass scale, then the point is assumed to reside in a neutral region.  Similarly to the excursion set model of halo statistics \cite{1974ApJ...187..425P,1991ApJ...379..440B,1993MNRAS.262..627L}, the mass function of ionized bubbles can be constructed by solving for the first-crossing distribution of random walks with an absorbing barrier obtained from eq.~(\ref{EQ:ESMRcondition}).  

In the ESMR, points in the IGM are assumed to be either fully ionized or fully neutral.  However, the average ionized fraction, $x_{\rm HII}(\eta,R),$ within a large spherical region of radius $R$ can be written as
\begin{equation}
x_{\rm HII}(\eta,R) = \xi_{\rm ESMR} f_{\rm coll}(\eta,M_{\rm min},R,\delta_{R})\,.
\end{equation} 
Hence, large-scale fluctuations in the ionized fraction simply correspond to large-scale fluctuations in the collapsed fraction.

There are four aspects of radiative transfer that distinguish the LPTR from the ESMR:  

(i) In ESMR, a collapsed object within a spherical region contributes to the ionization status of that region in proportion only to its mass, but not to its distance from the center. However, if a source is near the boundary of the sphere, not only is the flux of ionizing photons received at the center of the sphere reduced by the inverse-square of the distance, but a large fraction of ionizing photons from this source can leak out of the region, contributing to the ionization of neighboring regions.  So the number of atoms in a given region that are ionized by a source must depend on both the source mass and its separation from the center. 
On the other hand, let us recapitulate the basic principle of the LPTR; when the perturbations in the effective source emissivity are transferred to perturbations in the monopole radiation intensity, the mean optical depth is used to account approximately for the absorption of radiation along the ray.  Thus, the LPTR includes the basic elements of radiative transfer, albeit in an approximate way. 

(ii) The ESMR implicitly makes the ``infinite speed of light'' approximation, in which the light crossing time for ionizing photons is assumed to be much smaller than the time scale over which enough sources form to completely ionize a volume.  In this approximation, 
a region is assumed to be ionized instantaneously once it accumulates enough collapsed fraction. On the other hand, as is clear from our formulation above, the LPTR properly relates the arrival rate of photons to the emissivity elsewhere along the past light-cone (with the correct speed of light). The infinite speed of light approximation is reasonable for small H~II regions formed during the early stages of reionization. However, as H~II regions grow in size and the mean-free-path for ionizing photons grows, accounting for the finite speed of light becomes increasingly important. The distinction between finite and infinite speed of light is always important, in fact, for hard enough photons (e.g.\ X-rays), for which the mean-free-path is large from the outset.

(iii) In the ESMR, the global ionized fraction, $\bar{x}_{\rm HII}$, is the weighted sum of $x_{\rm HII}=1$ for fully ionized regions and $x_{\rm HII}=0$ for fully neutral regions; partially ionized regions are not produced in this model\footnote{Some semi-numerical simulation algorithms based on ESMR, e.g.\ Ref.~\cite{2011MNRAS.411..955M}, assign a partial ionization, $x_{\rm HII} = \xi_{\rm ESMR} f_{\rm coll}(\eta,M_{\rm min},R_{\rm cell},\delta_{R_{\rm cell}})$ (where $R_{\rm cell}$ is the cell size), to cells that do not meet the excursion set ionization criterion, eq. (\ref{EQ:ESMRcondition}). This ``partial'' ionization accounts for small H~II bubbles that are not resolved by the simulation, but the IGM is fundamentally modeled as  a two-phase medium.}.   A two-phase IGM is a reasonable approximation in scenarios where only UV sources reionize the universe.  In that case, the mean free path in the neutral IGM is extremely short, and the boundaries between H~II and neutral regions are sharp.  In contrast, the LPTR is not restricted to any single ionization topology. 

(iv)  Unlike the ESMR, the source spectrum appears explicitly in the LPTR formalism.  Together with (ii) and (iii), this feature makes the LPTR a better-suited analytical tool for investigating scenarios in which X-rays contribute significantly to reionization, and/or an earlier epoch of partial ionization and pre-heating of the IGM by the first X-ray sources.  

We note that the second and third points above are important only when the source spectrum contains significant portions of hard photons such as X-rays. While there has been no direct observational constraint for the spectrum hardness, 21~cm observations \cite{2014ApJ...788..106P,2015arXiv150300045P} have suggested that by $z\sim 7.7$, the IGM has been warmed from its cold primordial state. Since X-rays dominate the heating of the IGM, this suggests that the contribution from X-rays is likely to play a crucial, even though not dominant, role in reionization. 

The fundamental assumption of the LPTR on the linearization of radiative transfer may be a serious shortcoming. Is dropping $\Delta \Gamma_\nu \Delta I_\nu$ in eq.~(\ref{eqn:RTeqn5}) a good approximation? If the IGM is a two-phase medium with the ionized fraction taking the value of either zero or unity, this approximation breaks down on scales approaching the typical bubble size (about tens of comoving Mpc). However, on much larger scales, $k\lesssim 0.01\,{\rm Mpc}^{-1}$, the linearization scheme should be valid, at least at face value, because the fluctuations of ionized fraction are smoothed on those large scales. In comparison, the ESMR provides a non-perturbative model that is applicable on all scales, including the regime in which the LPTR breaks down.  

\section{Numerical Results}
\label{sec:numerical_results}

The strength of the LPTR lies in the fact that it can be applied ``right out of the box'' to arbitrary source models and recombination clumping factors, since it is derived directly from the equations of radiative transfer and ionization balance.  As we described above, the LPTR can even be applied to cases with hard source spectra (i.e. X-rays) to explore scenarios beyond the limits of the ESMR.  In this section, we demonstrate the wide range of models that can be readily covered by the LPTR, highlighting differences made by variations in the source and clumping factor models.  For reference, we also compare the ESMR to an LPTR model constructed to approximately match the assumptions of the ESMR. 
The numerical implementation in this section employs an LPTR code described in  Refs.~\cite{2007MNRAS.375..324Z,2013MNRAS.433.2900D}, which evaluates the LPTR solution in Fourier space. We refer readers to Refs.~\cite{2007MNRAS.375..324Z,2013MNRAS.433.2900D} for details on the Fourier space implementation techniques. 

\subsection{Illustrative models}

In the LPTR, one is free to write down a source model of any form.  In this paper we consider two, for illustrative purposes. In source Model A, the number of photons released during a time interval $d \eta$ is proportional to the {\it change} in the number of collapsed hydrogen atoms in that interval.  In this case, the emissivity can be written as \cite{2007MNRAS.375..324Z,2013MNRAS.433.2900D} 
\begin{equation}
S_\nu(\eta, {\bf x},f) = \frac{h_{\rm pl}}{4\pi} \gamma^{\rm A}(\nu) \frac{\partial}{\partial\eta} \left[ n_{\rm H}(\eta, {\bf x},R) f_{\rm coll}(\eta,M_{\rm min},R,\delta_R) \right]
\label{eqn:sourceA}
\end{equation}
where $n_{\rm H}(\eta, {\bf x},R)$ is the comoving number density of hydrogen atoms smoothed\footnote{In the large-scale limit $k\ll 2\pi/R$, i.e. the regime of interest here, the results are independent of the smoothing scale $R$.} over a spherical region of radius $R$, and $\gamma^{\rm A}(\nu)$ is the unitless source spectrum function. In other words, in this model, photon production is fueled only by {\it newly} collapsed hydrogen. Model A is similar to the default assumption of the ESMR; the total number of photons emitted from a region is proportional to the total collapsed mass within that region. In source Model B, the rate of photon production at a given time is proportional to the total number of collapsed hydrogen atoms \cite{2013MNRAS.433.2900D},
\begin{equation}
S_\nu(\eta, {\bf x},f) = \frac{h_{\rm pl}}{4\pi}H_0 \gamma^{\rm B}(\nu) n_{\rm H}(\eta, {\bf x},R) f_{\rm coll}(\eta,M_{\rm min},R,\delta_R)\,,
\label{eqn:sourceB}
\end{equation}
where the source spectrum, $\gamma^{\rm B}(\nu)$, corresponds to the number of ionizing photons per collapsed hydrogen atom per Hubble time $H_0^{-1}$ per unit $\ln(\nu/\nu_{\rm H})$.  In this model, photon production is continuously fueled by collapsed mass.  It is useful to compare source models A and B, not only because they represent different physical regimes, but also because the latter is often employed in radiative transfer simulations (see, e.g.\ \cite{2014MNRAS.439..725I}), whereas the former is the typical assumption in analytical and semi-numerical methods based on the ESMR. 

Note that LPTR, in either position- or Fourier-space, does not imply any
particular choice of the statistical relationship between the
density field and the distribution of ionizing sources. 
LPTR can be used with any sophisticated treatments of the source-density relationship, 
ranging from a local, linear halo bias model to something that correlates
the density field at different points in space and time (e.g.\ \cite{2005ApJ...624..491I}).

In both LPTR source models we consider herein, we evaluate the collapsed fraction according to the excursion-set approach, as in ESMR, 
\begin{equation}
f_{\rm coll}(\eta,M_{\rm min},R,\delta_R) = {\rm erfc}\left[ \frac{\delta_c - \delta_R(\eta)}{\sqrt{2[S_{\rm min}(\eta)-S_R(\eta)]}} \right]\,,
\end{equation}
where $\delta_c \approx 1.686$ is the critical extrapolated linear overdensity in the spherical collapse model in an Einstein-de Sitter universe, $S_R(\eta)$ is the variance of density fluctuations smoothed on scale $R$, and $S_{\rm min}(\eta)$ corresponds to the variance smoothed on the scale $R_{\rm min} = (3 M_{\rm min}/4\pi\bar{\rho}_m)^{1/3}$. 
As done in the original LPTR \cite{2007MNRAS.375..324Z,2013MNRAS.433.2900D}, we Taylor expand $f_{\rm coll}$ and $S_\nu$ to linear order in the filtered overdensity $\delta_R$ around $\delta_R = 0$, and take the limits of $S_R\to 0$ and $\tilde{\delta}_R({\bf k}) \approx \tilde{\delta}({\bf k})$, assuming $R$ is large enough, so that the bias of monopole radiation intensity with respect to the underlying density fluctuations can be analytically calculated in ${\bf k}$-space.

In both LPTR source models, we parameterize the source spectrum with a power law in $\nu$ \cite{2007MNRAS.375..324Z,2013MNRAS.433.2900D}, 
\begin{equation}
\gamma^{\rm A,B}(\nu) = \zeta^{\rm A,B}_{\rm LPTR} C_s\, (\nu/\nu_{\rm H})^{(1+s)}\,. 
\label{eqn:softness}
\end{equation}
The power-law index $s$ can be used to tilt the source spectrum to a soft one, in which UV photons are dominant, or a hard one, in which X-rays represent a larger fraction of photons. Here, we will use $s=-3$ and $s=-1$ as examples of soft and hard source spectra respectively. The normalization factor $C_s$ is defined such that integration over the spectrum always gives the efficiency parameter $\zeta^{\rm A,B}_{\rm LPTR}$, i.e.\ $\int_{\nu_{\rm H}}^{\infty} \gamma^{\rm A,B}(\nu) d\nu/\nu = \zeta^{\rm A,B}_{\rm LPTR}$.  In the case with $s=-3$, this leads to a normalization of $C_s=-(1+s)$.  On the other hand, the integral diverges for $s=-1$, so we smoothly truncate the spectrum at a cutoff frequency, $\nu_{\rm max}$.  In this case, we take the cutoff at $\ln(\nu_{\rm max}/\nu_{\rm H})=10$, or $h_{\rm pl}\nu_{\rm max}=300\,{\rm keV}$.  The values of the parameters $\zeta^{\rm A,B}_{\rm LPTR}$ are determined by fixing the electron elastic scattering optical depth to $\tau_{\rm es} = 0.08$, consistent with the {\it Planck} 1-year result \cite{2014A&amp;A...571A..16P}.  

Note that in the hard spectrum case, the helium photoionization rate should be taken into account in the radiative transfer and ionization balance equations, in principle. While the formulation given above in Sections~\ref{sec:reformulation} and \ref{sec:rederivation} is fully general for any photoionization species, we focus on the case in which only hydrogen reionization is considered in what follows for illustrative purposes, as in Ref.~\cite{2007MNRAS.375..324Z}. We refer readers to eqs.~(7) and (20) in Ref.~\cite{2007MNRAS.375..324Z} for solving the ionization balance equation for hydrogen. 

In addition to flexibility in source models, the LPTR can naturally accommodate any model for the time dependence of the clumping factor. In this paper, we consider two simple models of the recombination clumping factor, $C_{\rm HII} = \left< n_{\rm HII}^2 \right> /\left< n_{\rm HII} \right>^2$: one with constant $C_{\rm HII}=2$ for all time, where this value was chosen to fall within the range of plausible values suggested by numerical simulations (see, e.g.\ \cite{2012ApJ...747..100S,2012MNRAS.427.2464F}), and the other in which $C_{\rm HII}$ varies with redshift according to the fitting formula \cite{2007MNRAS.376..534I}, 
\begin{equation}
C_{\rm HII}=26.2917 \exp(-0.1822z+0.003505 z^2)\,.
\label{eqn:time-varying-clumping}
\end{equation}
The clumping factors for photoionization, $C^{(1)}_{\gamma {\rm H}}$ and $C^{(2)}_{\gamma {\rm H}}$, are also free parameters in the LPTR. They are defined as $C^{(1)}_{\gamma {\rm H}}=\left< n_{\rm HI} I_\nu \kappa\right>/\left< n_{\rm HI} \right> \left<I_\nu \right>\left<\kappa\right>$ and $C^{(2)}_{\gamma {\rm H}}=\left< n_{\rm HI} I_\nu \right>/\left< n_{\rm HI} \right> \left<I_\nu \right>$, where $\kappa(\nu,x_{\rm HII})$ accounts for local secondary reionization by fast photoelectrons produced when the energetic X-ray photons ionize hydrogen \cite{1985ApJ...298..268S}. For simplicity, we assume $C^{(1)}_{\gamma {\rm H}}=C^{(2)}_{\gamma {\rm H}}=1$ for all time for all of the LPTR models, which is a good approximation as suggested by hydrodynamical/radiative transfer simulations \cite{2007ApJ...657...15K,2010ApJ...724..244A}.

Note that the emission from hydrogen recombination to the ground state can be a source of reionization, too. Since the recombination emission is presumably isotropic at each point, the LPTR formalism can accommodate it in the source term, in principle. However, we neglect this emission in our demonstration of the LPTR models for simplicity, because it is not the driving force for reionization.

In Table~\ref{tab:models}, we summarize the different LPTR models\footnote{We note that the ESMR, LPTR-2As, -tAs and -tBs models are taken from Ref.~\cite{2013MNRAS.433.2900D}.} considered below.  As an example of the naming convention in Table~\ref{tab:models}, the model labeled LPTR-2As corresponds to $C_{\rm HII}=2$, source Model A, and a soft spectrum with $s=-3$.  We also consider an ESMR model with $\zeta_{\rm ESMR} = 50.2$, the value of which is chosen to yield $\tau_{\rm es} = 0.08$, like all of the LPTR models.  Since we consider only the simplest variant of the ESMR, which is limited to an effectively constant and homogenous recombination rate (degenerate with the $\zeta_{\rm ESMR}$ parameter), and to the assumption that the total number of photons emitted in a region is proportional to the local collapsed fraction (consistent with source Model A in the case of UV photons), our ESMR results are most comparable to the LPTR-2As model.  We note that Ref.~\cite{2005MNRAS.363.1031F,2013ApJ...771...35K,2014ApJ...787..146K} extended the ESMR to account for an inhomogeneous recombination rate and that it is, in principle, possible to modify the ESMR to be consistent with source model B.  However, we emphasize that the purpose of our comparison is two-fold: (1) to highlight the flexibility built into the LPTR by exploring a wide range of reionization models; and (2) to compare the ESMR to the LPTR in a case with similar assumptions. 

In what follows, we use a fiducial $\Lambda$CDM cosmology with parameters $\Omega_m=0.28$, $\Omega_\Lambda=0.72$, $\Omega_b=0.046$, $H_0=100h\,{\rm km}\,{\rm s}^{-1}\,{\rm Mpc}^{-1}$ with $h=0.7$, $n_s=0.96$ and $\sigma_8=0.82$, consistent with the {\it WMAP} seven-year result \cite{2011ApJS..192...18K} and the {\it Planck} 1-year result \cite{2014A&amp;A...571A..16P}. 

\begin{figure}
\includegraphics[width=0.5\textwidth]{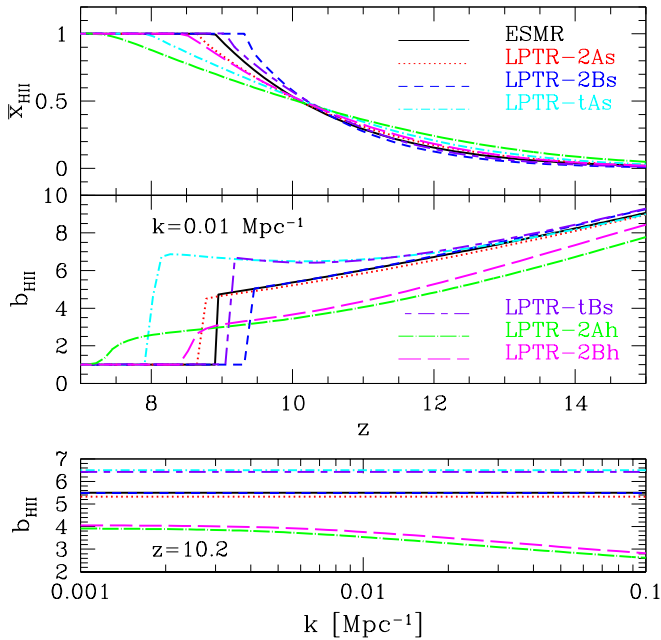}
\caption{Predictions of various models on reionization: (top) the global ionization fraction $\bar{x}_{\rm HII}$ as a function of redshift $z$, (middle) the ionized density bias $b_{\rm HII}(k,z)$ at $k=0.01\,{\rm Mpc}^{-1}$ as a function of redshift, and (bottom) $b_{\rm HII}(k,z)$ as a function of the wave number  $k$ at $z=10.2$ when $x_{\rm HII}\approx 0.5$. Shown are results of the ESMR (solid/black), \mbox{LPTR-2As} (dotted/red), \mbox{LPTR-2Bs} (short dashed/blue), \mbox{LPTR-tAs} (dot - short dashed/cyan), \mbox{LPTR-tBs} (short dash - long dashed/purple), \mbox{LPTR-2Ah} (dot - long dashed/green), \mbox{LPTR-2Bh} (long dashed/magenta). 
} 
\label{fig:comparison}
\end{figure}

\subsection{Results}

In Fig.~\ref{fig:comparison}, we compare two reionization observables across the models in Table~\ref{tab:models}: (top) the global ionized fraction $\bar{x}_{\rm HII}(z)$; (middle and bottom) the large-scale clustering of intergalactic H~II, quantified by the ionized density bias, $b_{\rm HII}(k,z)\equiv \tilde{\delta}_{\rho_{\rm HII}}({\bf k},z)/\tilde{\delta}_{\rho}({\bf k},z)$, where $\tilde{\delta}_{\rho_{\rm HII}}({\bf k},z)$ is the Fourier transform of the fractional H~II mass overdensity, $\delta_{\rho_{\rm HII}}({\bf x},z) \equiv \rho_{\rm HII}({\bf x},z)/ \bar{\rho}_{\rm HII} -1$, and $\delta_{\rho}({\bf k},z)$ corresponds to the fractional total matter overdensity. (For a more detailed discussion of the ionized density bias and its relation to other commonly used ionization bias parameters, see \S 2.1 of Ref. \cite{2013MNRAS.433.2900D}.) 

We find that the ESMR results agree, to a large extent, with those of the LPTR-2As model.   Recall that, among the LPTR models in Table~\ref{tab:models}, this model best reproduces the underlying assumptions of the ESMR; it uses source model A, a soft source spectrum dominated by UV photons, and a small, constant clumping factor, such that the effective recombination rate is not far from the recombination rate at the mean density of the universe. There is a small difference in the global ionization history at $\bar{x}_{\rm HII}>0.5$, and the end of reionization differs by $\Delta z \approx 0.3$ between these two models.  
These differences may be due to the fact that both the LPTR and the ESMR models considered here possess a deficiency in how they treat the end of reionization.\footnote{\label{MFPfootnote}In both of these models, the mean-free-path rapidly diverges at the end of reionization.  In reality, the mean-free-path cannot increase indefinitely due to absorption by Lyman-limit systems.} For example, the ``infinite speed of light'' assumption implicitly made in the ESMR breaks down near the end of reionization, and the assumption of linearized radiative transfer in LPTR becomes a bad approximation when H~II bubbles approach the large scale of interest as they grow.
On the other hand, the middle and bottom panels show that their ionized density biases are consistent with each other, aside from the timing difference at the end of reionization. 

Fig.~\ref{fig:comparison} shows that changing the clumping factor from a small, constant to the time-dependent model in eq.~(\ref{eqn:time-varying-clumping}), while holding the source model and $\tau_{\rm es}$ fixed, can slightly delay the end of reionization by $\Delta z \sim 0.3$ --- 0.6. In our time-dependent clumping model, the recombination clumping factor is significantly enhanced near the end of reionization, $C_{\rm HII}\approx 6.8$ at $z= 9$, more than 3 times larger than $C_{\rm HII}=2$.  
If the efficiency parameter were held fixed, then a larger clumping factor would always delay the end of reionization, which would yield a smaller $\tau_{\rm es}$. In order to match $\tau_{\rm es}$, therefore, we increased the efficiency parameter (see Table~\ref{tab:models}), so that the delay at the end of reionization is compensated by faster reionization at the beginning. Due to the larger source efficiency, a given source can ionize a larger volume by releasing more ionizing photons.
This significantly ($\lesssim 40\%$ fractional difference) enhances the ionized density bias. 

On the other hand, switching the type of source model can change the global ionization history significantly for $\bar{x}_{\rm HII}>0.5$. In particular, the end of reionization in source Model A is later than that in Model B by $\Delta z \sim 0.7$ --- 1, because ionizing photons are produced from all, not just new, collapsed mass in Model B. The ionized density bias, however, is almost unchanged, until the end of reionization. 

Finally, let us consider the impact of a source spectrum weighted towards X-rays -- a region of parameter space that the ESMR cannot explore.   In the hard-spectrum models (with $s=-1$), the end of reionization is delayed by $\Delta z \sim 0.8$ relative to their soft-spectrum counterparts.  The hard-spectrum models also produce a lower overall amplitude of $b_{\rm HII}$, a more-gradual decline of $b_{\rm HII}$ towards unity at the end of reionization, and a scale-dependence in $b_{\rm HII}(k,z)$, as shown in the bottom panel of Fig.~\ref{fig:comparison}.  All of these features can be explained in terms of the difference between the mean-free-paths of hard and soft photons.  In models with a soft source spectrum, H~II regions are bubble-like with sharp boundaries, and they begin around highly biased sources, which boosts the bias of H~II fluctuations.  The mean-free-path of UV photons remains much shorter than the length scales corresponding to $k=10^{-3}-0.1$ Mpc$^{-1}$ up until the very end of reionization. On scales much larger than the mean-free-path, the ionized density bias traces the source bias, which is assumed to be scale-independent in our calculations; hence, the ionized density bias is also scale-independent.   The mean-free-path rises very rapidly at the end of reionization as large H~II bubbles merge (see Ref. \cite{2013MNRAS.433.2900D}), rapidly suppressing the ionized density bias as reionization ends (see, however, footnote \ref{MFPfootnote}).  In contrast, since the long mean-free-path of hard photons results in a more homogeneously ionized IGM, the distribution of intergalactic H~II in the hard-spectrum models is not so highly clustered around biased sources, resulting in a lower overall amplitude of the ionized density bias.  The mean-free-path of X-rays can be as high as $\sim 100$ Mpc during reionization, corresponding to $k\sim 0.01$ Mpc$^{-1}$.  Since the bias of H~II fluctuations relative to the underlying density field is suppressed on scales below the mean-free-path, $b_{\rm HII}$ is suppressed above $k\gtrsim 0.01$ Mpc$^{-1}$, as shown in the bottom panel in Fig.~\ref{fig:comparison}.  Lastly, in the hard-spectrum models, the rise of the mean-free-path is not as rapid at the end of reionization as in the UV case, resulting in a more gradual suppression of $b_{\rm HII}$ towards unity.  

The above results demonstrate the built-in capability of the LPTR to explore a wide range of reionization scenarios, and to go beyond the ESMR in exploring scenarios involving X-rays.

\section{Conclusions}
\label{sec:conclusion}

We presented a new formulation of the LPTR that elucidates its approximation to radiative transfer in position space.  The fundamental approximation of the LPTR is that the propagation of perturbations in effective source emissivity is attenuated with a mean optical depth that is not only integrated along the past light-cone, but also averaged over all photon arrival directions and all points in space.  Our reformulation facilitates the interpretation of the LPTR and may inform future efforts to broaden its regime of applicability.  

We contrasted the underlying principles of the LPTR against those of another useful analytical model of reionization --- the ESMR. We also used an LPTR model constructed to match approximately the underlying assumptions of the ESMR to show that these two distinctive analytical approaches yield consistent results in this simple model. However, we further demonstrated that the LPTR can readily explore a wide range of parameter space in source models and recombination rate models. These physically-motivated assumptions in source emissivity help bridge the gap between ionizing sources at high redshift, which may be too faint to be directly observed in the foreseeable future, and radio observations of the EOR. More importantly, the LPTR is able to go beyond the limits of the ESMR in studying scenarios involving X-rays, which can play a leading role in partially reionizing and/or pre-heating the IGM during the early phases of reionization. 

Although the main emphasis of this paper has been the interpretation of the LPTR in position space, we note that, in practice, the LPTR is computationally fast and efficient due to its unique implementation of radiative transfer locally in Fourier space.  In fact, because of this, the LPTR can serve as the basis of a semi-numerical simulation algorithm\footnote{It should be clarified that while the equation of ionization balance is also linearized in the original LPTR formalism, it is unnecessary to do so in a semi-numerical approach, since it is straightforward to solve the ionization balance equation {\it locally} in position space. In other words, with a semi-numerical algorithm based on the LPTR, it is possible to incorporate fully nonlinear structures in the source and IGM density distributions into the ionization balance solver, while applying the linear approximation only to the radiative transfer solver.}, which can efficiently generate 3D realizations of H~II distributions, in a similar spirit to semi-numerical codes \cite{2009MNRAS.394..960C,2010MNRAS.406.2421S,2011MNRAS.411..955M,2014Natur.506..197F} based on the ESMR  (Mao et al. in prep.).   

Let us complete our discussion by commenting on the shortcomings of the LPTR.  We first point out that the ESMR can easily work out the H~II bubble size distribution \cite{2006MNRAS.365..115F} -- an important statistic of reionization -- while the LPTR cannot. 
Secondly, the assumption of linearized radiative transfer, which neglects the inhomogeneity of photoionization rates along light rays, breaks down on the scale of typical H~II bubbles. 
As will be shown in future work, the issue of H~II bubble size distributions can be addressed by the aforementioned semi-numerical algorithm based on the principles of LPTR.  The issue of inhomogeneous photoionization rates is a more serious deficiency of the LPTR. Clearly, this effect is most accurately addressed with 3D radiative transfer simulations, though at a significant computational cost. An alternative semi-analytical approach is to introduce effective linear terms in place of the higher-order term $\Delta\Gamma_\nu \Delta I_\nu$ which is missing in the LPTR, and calibrate those effective terms by numerical simulations. We leave it to future work to test this deficiency of the LPTR against full numerical simulations (Mao et. al. in prep.), and to implement the aforementioned effective theory approach.

\begin{acknowledgments}

We thank Zoltan Haiman, Lam Hui, Matt McQuinn, and the anonymous referee for helpful comments on this manuscript. 
This work has been done within the Labex ILP (reference ANR-10-LABX-63) part of the Idex SUPER, and received financial state aid managed by the Agence Nationale de la Recherche, as part of the programme Investissements d'avenir under the reference ANR-11-IDEX-0004-02. 
AD was supported by U.S.~NSF Grant No.~AST 1312724. 
BDW acknowledges funding through his Chaire d'Excellence from the Agence Nationale de la Recherche (ANR-10-CEXC-004-01). 
JZ was supported by the national science foundation of China under grant No.~11273018, the national basic research program of China (2013CB834900), the national Thousand Talents Program for distinguished young scholars, a grant(No.~11DZ2260700) from the Office of Science and Technology in Shanghai Municipal Government, and the T.~D.~Lee Scholarship from the High Energy Physics Center of Peking University. 
PRS was supported by U.S.~NSF Grants No.~AST-0708176 and No.~AST-1009799, and NASA Grants No.~NNX07AH09G and No.~NNX11AE09G. 
\end{acknowledgments}

\end{document}